\documentclass[12pt]{article}
\normalbaselines
\begin{document}

\newcommand{\nablab}{{\mathop
{\rule{0pt}{0pt}{\nabla}}\limits^{\bot}}\rule{0pt}{0pt}}

\begin{center}

{\bf DYNAMO-OPTICALLY ACTIVE MEDIA: \\
NEW ASPECTS OF THE MINKOWSKI-ABRAHAM CONTROVERSY}

\vspace{8mm}
{\bf Alexander B. Balakin and Timur Yu. Alpin}

\vspace{8mm}

{\it Department  of General Relativity and Gravitation, Institute of Physics \\
Kazan federal University, Kremlevskaya str. 18, Kazan 420008,
Russia

E-mail: Alexander.Balakin@kpfu.ru
E-mail: Timur.Alpin@kpfu.ru
}
\end{center}

\vspace{8mm}

\begin{abstract}

Based on the covariant variation formalism, two versions of the symmetric
effective stress-energy tensor of the electromagnetic field in a
dynamo-optically active relativistic media are reconstructed in the framework
of the tetrad and aether paradigms, respectively. We show that the energy
density scalars and pressure tensors coincide for both versions of the
stress-energy tensors, however, the corresponding energy flux four-vectors
happen to be different in general case. This mathematical fact adds new
arguments into the 100-year-long discussion, which is called Minkowski-Abraham
controversy and is connected with the correct definition of the
electromagnetic energy flux in a continuous media. We consider three examples:
first, the axionically active vacuum; second, the spatially isotropic moving
dielectric medium; third, the dynamo-optically active medium. We discuss
possible applications of the elaborated formalism.
\end{abstract}

\vspace{10mm}

\vspace{0.8cm}


\section{Introduction}

More than a century ago the term {\it Minkowski-Abraham controversy} appeared in
the scientific lexicon as the result of discussions of Minkowski
\cite{Minkowski}, Einstein and Laub \cite{Einstein}, and Abraham \cite{Abraham}.
These discussions were focused on the correct definition of the energy flux of
the electromagnetic field in continuous material media. The interest to this
problem was revived in 1950s - 1970s, in course of systematic elaboration of
covariant theory of electromagnetically active media (see, e.g., \cite{Moller}-
\cite{Bressan}). In the review \cite{Brevik1} Brevik formulated experimental
motivation of the interest to this problem, thus giving a new impetus to
investigations of the problem of electromagnetic energy transfer (see, e.g.,
\cite{Antoci1}-\cite{Brevik5} for the extension of discussions).

We attract attention of Readers to the problem of energy transfer in a Cosmic
Dark Fluid, which joins the Dark Energy and Dark Matter constituents and can be
considered as an electromagnetically active chiral medium
\cite{Symm}-\cite{B2018}.
The Dark Fluid is assumed to be electrically neutral, it does not contain
electrically charged particles, however, this cosmic substratum, being a
specific quasi-medium, can influence the electromagnetic field indirectly, and,
respectively, can contribute its own electrodynamic part into the total
stress-energy tensor of the Universe. One of the ways, which is open for the
Dark Fluid influence, is the so-called dynamo-optical activity of the moving
medium. This term was introduced in \cite{LLP} to describe polarization and
magnetization of a medium, which moves non-uniformly, i.e., when the medium flow
is characterized by the acceleration, shear, rotation and expansion.
When we deal with dynamo-optical interactions, we are faced with the problem how
to separate the dynamo-optical energy flow and the one of the
non-electromagnetic origin; in other words, we are faced again with the
classical alternative associated with the Minkowski-Abraham controversy. There
are at least three motives for studying the mentioned problem just now and
namely in this context.

The first motif is connected with the definition of the velocity four-vector,
which is the important player in the theory of the medium motion. On the one
hand, there is the classical  Landau-Lifshitz algebraic definition of the
velocity four - vector $V^i$, appeared as the time-like eigen-vector of the
medium stress-energy tensor; every cosmic constituent possesses such intrinsic
velocity. On the other hand, as an alternative, there exists a global unit
time-like vector field $U^i$, appeared in the Einstein-aether theory
\cite{J1}-\cite{J3}, which is associated with the velocity four-vector of some
quasi-medium, the dynamic aether. This global vector field defines the preferred
frame of reference \cite{N1,N2,N3}, thus providing the violation of the Lorentz
invariance of the theory \cite{LV1,LV2}. The model of dynamic aether is one of
the candidates for describing the Dark Energy phenomenon \cite{BSh}.

The second motif relates to the axionic extension of the cosmic electrodynamics,
which is associated with chirality of the cosmic medium. The pseudoscalar
(axion) field interacts with the electromagnetic field, with vector field
presenting the dynamic aether, and with gravitational field. When we study the
waves in the cosmic medium, we deal, in fact, not simply with pure
electromagnetic waves, but with a conglomerate of spin-0, spin-1 and spin-2
modes \cite{J2,B2018}. The corresponding cross-terms in the total stress-energy
tensor admit double interpretation, and we have to postulate: do they belong to
the electromagnetic part of the stress-energy tensor, or, e.g., to the part
associated with the axionic Dark Matter?

The third aspect is connected with the correct reconstruction of the
stress-energy tensor of the electromagnetic field. There exist the canonic and
effective stress-energy tensors of the system. The gravity field equations
operate with the symmetric effective stress-energy tensor, which can be
introduced using the variation procedure with respect to the space-time metric.
Since, independently of definition, the velocity four-vector is considered to be
normalized by unity, i.e., $g_{ik}V^i V^k =1$, or $g_{ik}U^i U^k =1$, this
vector quantity depends on metric and thus has to participate in the variational
procedure. Nevertheless, the variational procedures differ in the first and
second cases; in order to distinguish them we use later two terms: the {\it
tetrad paradigm}, and the {\it aether paradigm}, respectively. The first term
reflects the fact that when the velocity is the eigen-vector of the
stress-energy tensor, we can take it as the time-like unit vector $V^i =
X^i_{(0)}$ of the corresponding tetrad $\left\{X^i_{(a)} \right\}$.  The term
aether paradigm relates to the case, when the velocity four-vector is associated
with the unit time-like global vector field. In this context two questions
arise. The first question is: whether the whole effective stress-energy tensors
obtained by the variation procedure in the frameworks of the tetrad and aether
paradigms, coincide? The second question is typical for the Minkowski-Abraham
controversy: whether the electromagnetic energy flux vectors in the medium,
obtained in the tetrad and aether paradigms, coincide? Why the corresponding
difference can exist?

Also, we have to mention the following detail of discussion. The  energy flux
four-vector is known to appear as the result of application of the first or
second projection procedure to the stress-energy tensor of the electromagnetic
field (in the first procedure we project all the tensor quantities on the
direction $V^i$ and on hyper-surface orthogonal to it; in the second procedure
we use the four-vector $U^i$). However, in the tetrad paradigm the $V^i$
four-vector can be obtained as the eigen - vector either of the total
stress-energy tensor, or, e.g., as the one for its pure material constituent, or
for the Dark Fluid constituent. In other words, there exist an additional degree
of freedom for modeling of this four-vector. In the aether paradigm the unique
preferred global velocity four-vector plays this principal role, and there is no
additional variants for the choice.

To conclude, there is no {\it a priori} fixed answer for the question concerning
the structure and properties of the electromagnetic energy flux four-vector. The
goal of this work is to clarify the posed questions using the model of the
so-called dynamo-optical interactions in the framework of the
Einstein-Maxwell-aether-axion theory.

The paper is organized as follows. In Section II we recall the schemes of
derivation of the effective electromagnetic stress-energy tensors in the
framework of the tetrad and aether paradigms. In Section III we derive the
corresponding stress-energy tensors for the dynamo-optical interactions in the
chiral electrodynamic systems. Section IV contains the analysis of the following
three examples: the model of axionic vacuum, the model of spatially isotropic
homogeneous moving dielectric medium, and the model of dynamo-optically active
medium. We discuss the results in Section V.

\section{Basic formalism}

\subsection{Standard elements of the variation procedure}

The action functional of the theory, which we consider below, has the standard
structure:
\begin{equation}
S = \int d^4 x \sqrt{-g} \left\{ \frac{R+ 2\Lambda}{2\kappa}  + L_{({\rm
total})}\right\} \,, \label{lagran}
\end{equation}
where $g$ is the determinant of the metric, $R$ is the Ricci scalar, $\Lambda$
is the cosmological constant and $\kappa=\frac{8\pi G}{c^4}$ is the Einstein
constant. The Lagrangian of the physical system as a whole, $L_{({\rm total})}$,
can include the metric, pseudoscalar field $\phi$ and its gradient four-vector
$\nabla_k \phi$; it can contain vector field ($V^k$ or $U^i$ ) and the covariant
derivative ($\nabla_m V^k$ or $\nabla_m U^k$); the Maxwell tensor $F_{mn}$ also
can be the constructive element of the Lagrangian; finally, the Ricci and
Riemann tensors can appear, when one deals with the non-minimal version of the
theory (see, e.g., \cite{BL2005}).

The Einstein field equations appear as the result of variation with respect to
metric
\begin{equation}
R_{ik} - \frac{1}{2} g_{ik} R = \Lambda g_{ik} + \kappa
T_{ik}^{({\rm total})} \,, \label{Ein}
\end{equation}
where $R_{ik}$ is the Ricci tensor, and
the effective stress-energy tensor $T_{ik}^{({\rm total})}$ has the following
{\it formal} definition
\begin{equation}
T_{ik}^{({\rm total})} \equiv \frac{(-2)}{\sqrt{-g}} \frac{\delta}{\delta
g^{ik}} \left[\sqrt{-g} L_{({\rm total})}\right] \,. \label{2Ein}
\end{equation}
This tensor is symmetric by definition and has to be divergence-free due to the
Bianchi
identities:
\begin{equation}
T_{ik}^{({\rm total})} = T_{ki}^{({\rm total})} \,, \quad \nabla^k
T_{ik}^{({\rm total})} = 0 \,. \label{Tsymm}
\end{equation}
The total Lagrangian of the chiral dynamo-optically active system under
consideration can be reconstructed as the sum of four physically distinguished
parts
\begin{equation}
L_{({\rm total})} = L_{({\rm em})} + L_{({\rm ps})} + L_{({\rm vect})} +
L_{({\rm matter})} \,, \label{123}
\end{equation}
associated with the electromagnetic, pseudoscalar, vector fields and matter,
respectively.
Consider them in more detail.

\subsection{Master equations for the electromagnetic field }

We assume that the first (electromagnetic) part is quadratic in the Maxwell
tensor $F_{pq}$
\begin{equation}
L_{({\rm em})} = \frac14 C^{pqmn} F_{pq} F_{mn} \,, \label{em}
\end{equation}
and other parts of the Lagrangian do not contain the Maxwell tensor.
The Maxwell tensor is the anti-symmetrized derivative of the potential
four-vector $A_k$:
\begin{equation}
F_{mn} \equiv \nabla_m A_n - \nabla_n A_m  = \partial_m A_n - \partial_n A_m
\,.  \label{em130}
\end{equation}
The definition of the Maxwell tensor provides the first subset of master
equations of covariant electrodynamics
\begin{equation}
\nabla_l F_{mn} + \nabla_n F_{lm} +\nabla_m F_{nl} =0 \,,  \label{em13}
\end{equation}
which can be standardly rewritten in the compact form using the dual tensor
$F^{*ik}$:
\begin{equation}
F^{*ik} \equiv \frac12 \epsilon^{ikmn} F_{mn} \ \ \Rightarrow \nabla_k F^{*ik} =
0 \,.  \label{em14}
\end{equation}
Here $\epsilon^{ikmn}=\frac{E^{ikmn}}{\sqrt{-g}}$ is the Levi-Civita (pseudo)
tensor based on the absolutely skew-symmetric symbol $E^{ikmn}$ ($E^{0123}=1$).
The linear response tensor $C^{ikmn}$ possesses the evident symmetry of indices
\begin{equation}
C^{pqmn} = -C^{qpmn} = C^{mnpq} = -C^{pqnm} \,. \label{em2}
\end{equation}
We assume that the  tensor $C^{pqmn}$ can depend, first, on pseudoscalar field
$\phi$, second, on the vector field $V^i$ or $U^i$; third, linearly on the
gradient four-vector $\nabla_k \phi$, fourth,  linearly on the covariant
derivative $\nabla_k V^i$ or $\nabla_k U^i$. Such assumptions allow us to
describe the interactions between electromagnetic field and pseudoscalar field,
on the one hand, and  the coupling of the electromagnetic and vector fields.
Being the tensor quantity, $C^{pqmn}$ can include the metric, Kronecker deltas,
Levi-Civita tensor, as well as, the Riemann, Ricci tensors and Ricci scalar, if
one deals with the non-minimal theory.

The second subset of the master equations for the electromagnetic field can be
standardly obtained by variation of the action functional with respect to the
potential four-(co)vector $A_i$. This procedure yields
\begin{equation}
\nabla_k \left[C^{ikmn} F_{mn}\right]= - \frac{4\pi}{c} J^i \,, \label{em11}
\end{equation}
where the four-vector $J^i$ is the  electric current defined formally as
\begin{equation}
J^i \equiv \frac{1}{4\pi} \frac{\delta L_{(\rm matter})}{\delta A_i}\,.
\label{em12j}
\end{equation}
It is convenient to use the skew-symmetric induction tensor $H^{ik}$ defined as
\begin{equation}
H^{ik} \equiv  C^{ikmn} F_{mn} \,,  \label{em12}
\end{equation}
which is the divergence - free one, when the medium is non-conducting, i.e.,
$J^i=0$.

\subsection{Master equation for the pseudoscalar field}

We assume that the second (pseudoscalar) part of the Lagrangian is quadratic in
the gradient four-vector $\nabla_k \phi$, and contain the dimensionless
pseudoscalar field $\phi$ in even combinations
\begin{equation}
L_{({\rm ps})} = \frac12 \Psi^2_0 \left[ - {\cal C}^{mn} \nabla_m \phi \nabla_n
\phi + {\cal V}(\phi^2)\right] \,. \label{emps}
\end{equation}
The constitutive tensor ${\cal C}^{mn}$ is assumed to depend on the metric,
Kronecker deltas, Levi-Civita (pseudo) tensor,  and on the velocity and its
covariant derivative.
${\cal V}(\phi^2)$ is the potential of the pseudoscalar field; the parameter
$\Psi_0$ is reciprocal to the axion-photon coupling constant
$\frac{1}{\Psi_0}=g_{A \gamma \gamma}$.
Master equations for the pseudoscalar field have the form
\begin{equation}
\nabla_m \left[{\cal C}^{mn} \nabla_n \phi \right] + \phi {\cal
V}^{\prime}(\phi^2) = {\cal J} \,,
\label{ax1}
\end{equation}
where the pseudoscalar source is explicitly quadratic in the Maxwell tensor
\begin{equation}
{\cal J} = - \frac{1}{4 \Psi^2_0} F_{pq} F_{mn}\frac{\partial}{\partial \phi}
C^{pqmn}  +  \frac{1}{4 \Psi^2_0} \nabla_j \left[F_{pq}
F_{mn}\frac{\partial}{\partial(\nabla_j \phi)} C^{pqmn} \right] \,,
\label{ax2}
\end{equation}
and can depend on the vector field and its covariant derivative, when the linear
response tensor $C^{pqmn}$ is correspondingly extended.

\subsection{Master equations for the vector field: I. The tetrad paradigm}

The tetrad paradigm assumes that there is no additional part in the Lagrangian,
i.e., $L_{(\rm vect)} = 0$, and the velocity four-vector $V^i$ is the
eigen-vector of the effective stress-energy tensor $T_{ik}^{({\rm total})}$:
\begin{equation}
T_{ik}^{({\rm total})} V^k  = W_{(\rm total)} V_i  \,. \label{eigen1}
\end{equation}
The vector $V^i$ is assumed to be time-like and unit
\begin{equation}
g_{ik} V^i V^k = 1 \,, \label{eigen2}
\end{equation}
so that the corresponding eigen-value $W_{(\rm total)}$
\begin{equation}
W_{(\rm total)} = V^i T_{ik}^{({\rm total})} V^k   \label{eigen3}
\end{equation}
can be indicated as the energy density scalar. With this definition (it is
usually indicated as the Landau-Lifshitz definition)
the structure of the effective stress-energy tensor is
\begin{equation}
T_{ik}^{({\rm total})} =  W_{(\rm total)} V_i V_k  + {\cal P}_{ik}^{(\rm total)}
\,. \label{T99}
\end{equation}
Here the  tensor ${\cal P}_{ik}^{(\rm total)}$ is symmetric, orthogonal to the
velocity $V^i$ and describes the total pressure tensor of the system.
In this approach the velocity four-vector has to satisfy the master equations,
which are derived from the conservation law (\ref{Tsymm}). Indeed, the
divergence of of the tensor (\ref{T99}) is equal to zero, when
\begin{equation}
V^k \nabla_k [W_{(\rm total)} V_i] + W_{(\rm total)} V_i (\nabla_k V^k)  +
\nabla^k {\cal P}_{ik}^{(\rm total)} = 0 \,. \label{Tetrad1}
\end{equation}
As usual, the projection of (\ref{Tetrad1}) on the direction pointed by the
velocity $V^i$ gives the equation of the energy density  evolution
\begin{equation}
D W_{(\rm total)} + W_{(\rm total)}\Theta  =  {\cal P}_{ik}^{(\rm
total)}\nabla^k V^i \,, \label{Tetrad2}
\end{equation}
where $D \equiv V^k \nabla_k$ is the convective derivative, and $\Theta  {=}
\nabla_k V^k$ is the extension scalar of the velocity field. The projection of
(\ref{Tetrad1}) on the hyper-surface orthogonal to the velocity four-vector
yields
\begin{equation}
W_{(\rm total)} D V^s   + \Delta^{is} \nabla^k {\cal P}_{ik}^{(\rm total)} = 0
\,, \label{Tetrad3}
\end{equation}
where $\Delta^{is} \equiv g^{is}-V^i V^s$ is the projector, which is known to
possess  the following properties:
\begin{equation}
\Delta^{is} = \Delta^{si} \,, \quad \Delta^{is}V_s =0 \,, \quad \Delta^{s}_s =3
\,, \quad  \Delta^{is}\Delta_{js} = \Delta^{i}_j \,. \label{Tetrad4}
\end{equation}
Thus, the unit time-like velocity four-vector $V^i$ in the tetrad paradigm has
to satisfy the equations (\ref{Tetrad3}).

The  velocity four-vector $V^i$ can be included into the set of tetrad vectors
$X^i_{(a)}$; the index $(a)$ takes the values $(0),(1),(2),(3)$, and
$X^i_{(0)} \equiv V^i $. This quartet of four-vectors satisfies the
orthogonality - normalization conditions
\begin{equation}
g_{ik} X^i_{(a)}X^k_{(b)} = \eta_{(a)(b)} \,, \label{tetra1}
\end{equation}
\begin{equation}
\eta^{(a)(b)} X^p_{(a)}X^q_{(b)} = g^{pq} \,, \label{tetra111}
\end{equation}
where $\eta_{(a)(b)}$ denotes the Minkowski matrix, diagonal
$(1,{-}1,{-}1,{-}1)$. Clearly, the tetrad four-vectors are linked by
the relation containing the metric, thus, we have to define the working formulas
for the variation $\frac{\delta
X^j_{(a)}}{\delta g^{ik}}$. This procedure is described in \cite{B2007}, we
recall the main details of this procedure.
First, the variation of (\ref{tetra111}) yields
\begin{equation}
\delta g^{pq} = \eta^{(c)(d)} \left[ X^q_{(d)} \delta X^p_{(c)} +
X^p_{(c)} \delta X^q_{(d)} \right]  \,, \label{tetra3}
\end{equation}
thus, we obtain the consequence
\begin{equation}
X_p^{(a)}\delta g^{pq}X_q^{(b)} =  \left[ X_p^{(a)} \delta X^p_{(c)}
\eta^{(c)(b)}+
\delta X^q_{(d)}X_q^{(b)} \eta^{(a)(d)} \right]  \,. \label{tetra31}
\end{equation}
Second, the variation $\delta X^i_{(a)}$ can be decomposed as the linear
combination  of the tetrad  four-vectors:
\begin{equation}
\delta X^i_{(a)} = X^i_{(f)} Y^{ \ \ (f)}_{(a)} \,. \label{tetra5}
\end{equation}
If we put (\ref{tetra5}) into (\ref{tetra31}) we obtain
\begin{equation}
Y^{(a)(b)} + Y^{(b)(a)} = \delta g^{pq} X_{p}^{(a)}X_{q}^{(b)} \,.
\label{tetra6}
\end{equation}
Generally, the object $Y^{(a)(b)}$ has the symmetric and antisymmetric parts,
$Y^{(a)(b)}= Y^{((a)(b))}+ Y^{[(a)(b)]}$, however, only the symmetric part is
assumed to be formed by the metric variation; this idea gives immediately that
\begin{equation}
\delta X^i_{(a)} = \frac{1}{4} \delta g^{pq} \left[X_{p (a)}
\delta^i_q + X_{q (a)} \delta^i_p \right]  \,, \label{tetra8}
\end{equation}
and consequently, for the velocity four-vector we have
\begin{equation}
\frac{\delta V^i}{\delta g^{pq}} = \frac{1}{4}  \left[V_{p} \delta^i_q + V_{q}
\delta^i_p \right] \,,
\quad \frac{\delta V_i}{\delta g^{pq}} = - \frac{1}{4}  \left[V_{p} g_{iq} +
V_{q} g_{ip} \right]  \,. \label{tetra9}
\end{equation}
When the linear response tensor $C^{ikmn}$ depends on the covariant derivative
of the velocity four-vector, we need to prepare the formula for variation of
$\nabla_m V^l$:
$$
\delta[\nabla_m V^l] = \nabla_m (\delta V^l) + V^n \delta \Gamma^l_{mn} =
$$
\begin{equation}
=\frac14 \delta g^{pq}\left(\delta^l_p \nabla_m V_q + \delta^l_q \nabla_m V_p
\right) + \frac14 \left(V_p g_{mq}+V_q g_{mp} \right) \nabla^l \delta g^{pq} -
\frac14 \left(\delta^l_p g_{mq}+\delta^l_q g_{mp} \right)V^n \nabla_n \delta
g^{pq} \,. \label{tetra10}
\end{equation}
Clearly, it contains the terms of the type $\nabla_n \delta g^{pq}$, and thus
the variation procedure requires the corresponding integration by part, when we
calculate the stress-energy tensor of the electromagnetic field.

\subsection{Master equations for the vector field: II. The aether paradigm}

The aether paradigm assumes that there exist an additional  time-like vector
field $U^i$, and it has to be included into variation procedure as an
independent player. To be more precise, the corresponding part of the Lagrangian
is non-vanishing
\begin{equation}
L_{({\rm vect})} = \frac{1}{2\kappa} \left[\lambda(g_{pq}U^pU^q -1) + K^{ab}_{\
\ mn} \nabla_a U^m \nabla_b U^n \right] \,, \label{v1}
\end{equation}
the function $\lambda$ is the Lagrange multiplier providing the vector field to
be normalized by unity; the Jacobson's constitutive tensor $K^{ab}_{\ \ mn}$ is
of the form
\begin{equation}
K^{ab}_{\ \ mn} = C_1 g^{ab} g_{mn} + C_2 \delta^a_m \delta^b_n + C_3 \delta^a_n
\delta^b_m  + C_4 U^a U^b g_{mn} \,, \label{v2}
\end{equation}
where $C_1$, $C_2$, $C_3$ and $C_4$ are the phenomenological parameters (see,
e.g., \cite{J1}). The term (\ref{v1}) is the participant of three variation
procedures. First, the variation with respect to the Lagrange multiplier
$\lambda$ yields $g_{pq}U^pU^q =1$,
i.e., the vector field is normalized by unity, and thus it is time-like
everywhere; these properties support the idea to consider this vector field as
the one of a global velocity.
Second, the variation of the total action functional with respect to the vector
field $U^i$ gives following equation:
$$
\frac{\lambda}{\kappa} U_j - \frac{1}{\kappa}\nabla_a \left[K^{ab}_{\ \ jn}
\nabla_b U^n \right] + \frac{1}{\kappa} C_4 \nabla_j U^n U^b \nabla_b U_n +
\frac14 F_{pq}F_{mn} \frac{\partial C^{pqmn}}{\partial U^j}
-\frac14 \nabla_l \left[F_{pq}F_{mn} \frac{\partial C^{pqmn}}{\partial (\nabla_l
U^j)} \right] -
$$
\begin{equation}
-\frac12 \Psi^2_0 \nabla_m \phi \nabla_n \phi \ \frac{\partial C^{mn}}{\partial
U^j}
+ \frac12 \Psi^2_0 \nabla_l \left[\nabla_m \phi \nabla_n \phi \ \frac{\partial
C^{mn}}{\partial (\nabla_l U^j)} \right] =0   \,.
\label{vec31}
\end{equation}
This equation can be rewritten in the well-known form
\begin{equation}
\nabla_a {\cal J}^{a}_{\ j} =  I_j^{(\rm U)} + \kappa I_j^{(\rm F)} + \kappa
I_j^{(\phi)} + \lambda \ U_j  \,,
\label{vec32}
\end{equation}
where the following definitions are used:
\begin{equation}
{\cal J}^{a}_{\ j} = K^{ab}_{\ \ jn} (\nabla_b U^n) \,, \quad I_j^{(\rm U)} =
C_4 \nabla_j U^n U^b \nabla_b U_n \,,
\label{J0}
\end{equation}
\begin{equation}
I_j^{(\rm F)} = \frac14 F_{pq}F_{mn} \ \frac{\partial C^{pqmn}}{\partial U^j}
-\frac14 \nabla_l \left[F_{pq}F_{mn} \ \frac{\partial C^{pqmn}}{\partial
(\nabla_l U^j)} \right] \,,
\label{J1}
\end{equation}
\begin{equation}
I_j^{(\phi)} = -\frac12 \Psi^2_0 \nabla_m \phi \nabla_n \phi \ \frac{\partial
C^{mn}}{\partial U^j} + \frac12 \Psi^2_0 \nabla_l \left[\nabla_m \phi \nabla_n
\phi  \ \frac{\partial C^{mn}}{\partial (\nabla_l U^j)} \right] \,.
\label{J2}
\end{equation}
Clearly, using the projection of the equation (\ref{vec32}) on the direction
$U^j$ and the normalization condition we can obtain the Lagrange multiplier
\begin{equation}
\lambda = U^j \left[\nabla_a {\cal J}^{a}_{\ j} -   I_j^{(\rm U)} - \kappa
I_j^{(\rm F)} - \kappa I_j^{(\phi)} \right]  \,.
\label{vec33}
\end{equation}
As well, using the projector $\Delta^{ik}=g^{ik}-U^iU^k$, we can obtain the
equation
\begin{equation}
\Delta^{sj} \nabla_a {\cal J}^{a}_{\ j} = \Delta^{sj} \left[I_j^{(\rm U)} +
\kappa I_j^{(\rm F)} + \kappa I_j^{(\phi)}\right] \,,
\label{vec35}
\end{equation}
which includes the velocity four-vector but does not contain the Lagrange
multiplier.

\subsection{Standard auxiliary tensor quantities and their interpretation}

\subsubsection{Decomposition of the covariant derivative of the velocity
four-vector}

The covariant derivative $\nabla_k$ is known to be presented as the
decomposition on the longitudinal and transversal components with respect to
chosen velocity four-vector; when we deal with the vector field $U^i$, we have,
respectively:
\begin{equation}
\nabla_k = U_i DU_k + \nablab_k \,,  \quad D \equiv U^m \nabla_m  \,, \quad
\nablab_k \equiv \Delta_k^m \nabla_m \,, \quad \Delta_k^m \equiv \delta^m_k -U^m
U_k \,,
\label{act30}
\end{equation}
where $D$ is the convective derivative, and $\Delta_k^m$ is the projector. In
these terms the tensor $\nabla_i U_k$ can be represented as follows:
\begin{equation}
\nabla_i U_k = U_i DU_k + \sigma_{ik} + \omega_{ik} +
\frac{1}{3} \Delta_{ik} \Theta \,, \label{act3}
\end{equation}
where $DU^{i}$ is the acceleration four-vector, $\sigma_{ik}$ is the symmetric
trace-free shear tensor, $\omega_{ik}$ is
the skew-symmetric vorticity tensor, and $\Theta$ is the expansion scalar. The
definitions of these quantities are well-known
$$
DU_k \equiv U^m \nabla_m U_k \,,
\quad
\sigma_{ik}
\equiv \frac{1}{2} \left(\nablab_i U_k {+}
\nablab_k U_i \right) {-} \frac{1}{3}\Delta_{ik} \Theta  \,,
$$
\begin{equation}
\omega_{ik} \equiv \frac{1}{2} \left(\nablab_i U_k {-} \nablab_k U_i \right)
\,,
\quad
 \Theta \equiv \nabla_m U^m = \nablab_m U^m  \,.
\label{act4}
\end{equation}
The terms acceleration, shear, vorticity and expansion relate in this case to
the aether flow. When we deal with the velocity four-vector $V^i$, the
decomposition is similar.

\subsubsection{Decomposition of the Maxwell tensor $F_{ik}$ and of the induction
tensor $H^{mn}$}

Electrodynamics of continuous media operates with the quartet of four-vectors
$D^i$, $E^i$, $H^i$ and $B^i$. When one deals with the velocity four-vector
$V^k$, these quantities are defined as follows:
\begin{equation}
D^i \equiv H^{ik}V_k \,, \quad H_i \equiv H^{*}_{ik}V^k \,,
\quad E^i \equiv F^{ik} V_k \,, \quad B_i \equiv F^{*}_{ik} V^k \,.
\label{DHEB}
\end{equation}
When we work in the aether paradigm, we have to replace $V^k$ with $U^k$. The
four-vectors
$D^i$, $E^i$, $H^i$ and $B^i$ are orthogonal to the corresponding velocity
four-vector. In these terms the tensors
$F_{ik}$, $F^{*}_{ik}$, $H^{ik}$ and $H^{*ik}$ can be represented as follows:
\begin{equation}
F_{ik} = E_i V_k - E_k V_i - \epsilon_{ikmn}B^m V^n  \,,
\quad
F^*_{ik} = B_i V_k - B_k V_i + \epsilon_{ikmn}E^m V^n \,,
\label{FEBHDH}
\end{equation}
\begin{equation}
H^{ik} = D^i V^k - D^k V^i - \epsilon^{ikmn}H_m V_n \,,
\quad
H^{*ik} = H^i V^k - H^k V^i + \epsilon^{ikmn}D_m V_n \,. \label{6FEBHDH}
\end{equation}
$E^i$ can be interpreted as the four-vector of electric field found in the frame
of reference associated with the velocity four-vector $V^m$. $B^i$ describes the
magnetic induction, $D^i$ corresponds to the electric induction, $H_i$ can be
indicated as the four-vector of the magnetic field.

\subsubsection{Decomposition of the linear response tensor}

The tensor $C^{ikmn}$ symmetric with respect to the pair index transposition
$C^{mnik}=C^{ikmn}$, also can be decomposed using the appropriate vector field;
when we deal with the four-vector $V^k$ the corresponding decomposition is (see,
e.g., \cite{1,2} for details):
$$
C^{ikmn} = \frac12 \left[ \varepsilon^{im} V^k V^n -
\varepsilon^{in} V^k V^m + \varepsilon^{kn} V^i V^m - \varepsilon^{km} V^i V^n
\right] -
$$
\begin{equation}
-\frac12 \eta^{ikl}(\mu^{-1})_{ls}  \eta^{mns} {-}
\frac12 \left[\eta^{ikl}(V^m \nu_{l}^{ \ n} {-} V^n \nu_{l }^{ \
m}) {+} \eta^{lmn}(V^i \nu_{l}^{ \ k} {-} V^k \nu_{l}^{ \ i} )
\right] \,. \label{Cdecomp}
\end{equation}
The new two-indices tensors are defined as follows:
\begin{equation}
\varepsilon^{im} = 2 C^{ikmn} V_k V_n\ , \quad (\mu^{-1})_{pq} = -
\frac{1}{2} \eta_{pik} C^{ikmn} \eta_{mnq}\ ,
\quad \nu_{p}^{ \ m} = \eta_{pik} C^{ikmn} V_n \,, \label{emunu}
\end{equation}
where $\eta_{pik} \equiv \epsilon_{pikq}V^q$.  The tensors $\varepsilon_{ik}$
and $(\mu^{-1})_{ik}$ are
symmetric, $\nu_{l}^{ \ k}$ is, in general, non-symmetric; they are
orthogonal to $V^i$, i.e.
\begin{equation}
\varepsilon_{ik} V^k = 0 \,, \quad  (\mu^{-1})_{ik} V^k = 0 \,, \quad
 \nu_{l}^{ \ k} V^l = 0 = \nu_{l}^{ \ k} V_k \,.
\label{ortho}
\end{equation}
The tensor $\varepsilon_{ik}$ is interpreted as the dielectric permeability
tensor
found in the frame of reference associated with the velocity four-vector $V^i$;
the tensor $(\mu^{-1})_{ik}$ describes the magnetic impermeability of the
medium; the  tensor $\nu_i^{ \ k}$ contains the so-called magneto-electric
coefficients of the medium.
This interpretation is based on the formula
\begin{equation}
D^i = \epsilon^{ik} E_k - \nu^{ \ i}_{k} B^k \,, \quad H_i = \nu^{
\ k}_{i} E_k + (\mu^{-1})_{ik} B^k \,, \label{DEBHEB}
\end{equation}
which can be directly obtained using the definitions presented above.

\section{Effective stress-energy tensor of the electromagnetic field in a
dynamo-optically active medium}

A number of details of the variation formalism based on the tetrad and aether
paradigms coincide. For instance, the following auxiliary  variational
identities are of common use:
\begin{equation}
\frac{\delta }{\delta g^{ik}} \phi = 0 \,, \quad \frac{\delta }{\delta g^{ik}}
\nabla_m \phi = 0 \,,
\quad \frac{\delta }{\delta g^{ik}} F_{mn} = 0 \,, \quad
\frac{1}{\sqrt{-g}}\frac{\delta \sqrt{-g}}{\delta g^{ik}} =
-\frac12 g_{ik} \,,
\label{em8}
\end{equation}
\begin{equation}
\frac{\delta g_{ls}}{\delta g^{ik}} = -
\frac12 \left[g_{li} g_{ks} + g_{lk} g_{is} \right] \,,  \quad  \frac{\delta
}{\delta g^{ik}} \delta_{p}^{q} = 0 \,,  \quad
\frac{\delta \epsilon^{lsrt}}{\delta g^{ik}} = \frac12 \epsilon^{lsrt} g_{ik}
\,.
\label{em9}
\end{equation}
However, all the details of procedures, which relate to variation with respect
to velocity four-vector  and its covariant derivative, have to be considered
individually, if we follow tetrad or aether paradigms.

\subsection{Calculations in the framework of the tetrad paradigm}

In the framework of the tetrad formalism we use the following definition of the
electromagnetic stress-energy tensor:
\begin{equation}
T_{ik}^{({\rm em})} = - \frac{1}{2\sqrt{-g}} \frac{\delta}{\delta g^{ik}}
\left[\sqrt{-g} F_{pq} F_{mn} C^{pqmn}\right] \,. \label{em4}
\end{equation}
Taking into account (\ref{em8}) we obtain immediately
\begin{equation}
T_{ik}^{({\rm em})} = \frac14 F_{pq} F_{mn} C^{pqmn}-
\frac12 F_{pq} F_{mn} \frac{\delta}{\delta g^{ik}} C^{pqmn}
\,. \label{em49}
\end{equation}
The first term is the scalar $\frac14 H^{mn} F_{mn}$, which is the part of all
known stress-energy tensors of the electromagnetic field in media; the
difference between them appears due to the second term. Keeping in mind
(\ref{em9}) we can rewrite (\ref{em49}) as follows:
$$
T_{ik}^{({\rm em})} =  \frac14 F_{pq} F_{mn} \left\{g_{ik} \left[C^{pqmn}-
\epsilon^{lsrt} \ \frac{\partial C^{pqmn}}{\partial \epsilon^{lsrt}} \right]-
(\delta^l_i \delta^s_k + \delta^l_k \delta^s_i) \ \frac{\partial
C^{pqmn}}{\partial g^{ls}}  - \right.
$$
$$
\left. - \frac12 \left(V_i \delta^j_k + V_k \delta^j_i \right) \ \frac{\partial
C^{pqmn}}{\partial V^j} - \frac12 \left(\delta^j_i \nabla_l V_k + \delta^j_k
\nabla_l V_i \right) \ \frac{\partial C^{pqmn}}{\partial (\nabla_l V^j)}
\right\} +
$$
\begin{equation}
+ \frac18 \nabla^j \left[\left(V_i g_{lk}+ V_k g_{li}\right)F_{pq} F_{mn}
\frac{\partial C^{pqmn}}{\partial (\nabla_l V^j)} \right] -
\frac18 \left(\delta^j_i g_{lk}+ \delta^j_k g_{li}\right)\nabla_s \left[F_{pq}
F_{mn} V^s \frac{\partial C^{pqmn}}{\partial (\nabla_l V^j)} \right]
\,. \label{em5}
\end{equation}
One has to stress that we deal with the example of the theory, in which the
stress-energy tensor contains not only the Maxwell tensor, but its covariant
derivative $\nabla_s F_{mn}$ also, since the linear response tensor $C^{pqmn}$
is assumed to contain the dynamo-optical terms, i.e., since  $\frac{\partial
C^{pqmn}}{\partial (\nabla_l V^j)} \neq 0$.

\subsection{Calculations in the framework of the aether paradigm}

Now we consider the vector field $U^j$ to be independent on the variation of the
metric, i.e., in contrast to (\ref{tetra9}), we have $\frac{\delta U^j}{\delta
g^{ik}} =0$. Also, we keep in mind, that the variation of the term $\frac12
\lambda (g_{mn}U^m U^n -1)$ with respect to metric $g^{ik}$ gives the
contribution $\lambda U_i U_k$ into the total stress-energy tensor. The quantity
$\lambda$ given by (\ref{vec33}) contains the part $I_j^{(\rm F)}$, which
according to (\ref{J1}) is quadratic in the Maxwell tensor; we add this term to
the stress-energy tensor of the electromagnetic field. The variation of the
covariant derivative also differ from (\ref{tetra10}), being of the following
form:
\begin{equation}
\delta (\nabla_l U^j) = - \frac12 \left[\delta^j_{(i} g_{k)l} U^n \nabla_n +
\delta^j_{(i} U_{k)}  \nabla_l - U_{(i} g_{k)l}  \nabla^j \right]\delta g^{ik}
\,. \label{em67}
\end{equation}
We use here and below the standard definition of the symmetrization:
$A_{(i}B_{k)} \equiv \frac12 (A_iB_k{+}A_kB_i)$.
Now the stress-energy tensor of the electromagnetic field can be written in the
following form
$$
{\cal T}_{ik}^{({\rm em})} =  \frac14 F_{pq} F_{mn} \left\{g_{ik}
\left[C^{pqmn}-
\epsilon^{lsrt} \ \frac{\partial C^{pqmn}}{\partial \epsilon^{lsrt}} \right]-
(\delta^l_i \delta^s_k + \delta^l_k \delta^s_i) \ \frac{\partial
C^{pqmn}}{\partial g^{ls}}  \right\} +
$$
$$
 - \frac14 U_iU_k U^j \left\{ F_{pq}F_{mn} \ \frac{\partial C^{pqmn}}{\partial
 U^j}
- \nabla_l \left[F_{pq}F_{mn} \ \frac{\partial C^{pqmn}}{\partial (\nabla_l
U^j)} \right] \right\}
$$
$$
 - \frac18  \nabla_l \left[\left(\delta^j_i U_k  + \delta^j_k U_i \right) \
 F_{pq} F_{mn} \frac{\partial C^{pqmn}}{\partial (\nabla_l U^j)} \right] +
$$
\begin{equation}
+ \frac18 \nabla^j \left[\left(U_i g_{lk}+ U_k g_{li}\right)F_{pq} F_{mn}
\frac{\partial C^{pqmn}}{\partial (\nabla_l U^j)} \right] -
\frac18 \left(\delta^j_i g_{lk}+ \delta^j_k g_{li}\right)\nabla_n \left[F_{pq}
F_{mn} U^n \frac{\partial C^{pqmn}}{\partial (\nabla_l U^j)} \right]
\,. \label{em52}
\end{equation}
Let us recall how to reconstruct the basic (irreducible) elements of the
stress-energy tensor of the electromagnetic field.

\subsection{Energy density, energy flux four-vector and the pressure tensor of
the electromagnetic field: Do they differ in the tetrad and aether paradigms?}

The standard decomposition of the symmetric effective stress-energy tensor
contains three basic elements: the energy density scalar $W$, the flux
four-vector ${\cal Q}^k$ and the pressure tensor ${\cal P}^{ik}$. In the
framework of the tetrad paradigm they are defined, respectively, as
\begin{equation}
W  \equiv V^m T_{mn}^{({\rm em})} V^n \,,
\label{decT1}
\end{equation}
\begin{equation}
{\cal Q}^k  \equiv V^m T_{mn}^{({\rm em})} \Delta^{kn} = \Delta^{km}
T_{mn}^{({\rm em})} V^n  \,, \label{decT2}
\end{equation}
\begin{equation}
{\cal P}^{ik}  \equiv \Delta^{im} T_{mn}^{({\rm em})} \Delta^{kn} \,.
\label{decT3}
\end{equation}
In order to obtain the corresponding quantities in the framework of the aether
paradigm we have to replace $U^j$ with $V^j$ and $T_{mn}^{({\rm em})}$ with
${\cal T}_{mn}^{({\rm em})}$. We are interested to calculate the difference
\begin{equation}
\tau_{ik} \equiv {\cal T}_{ik}^{({\rm em})}- T_{ik}^{({\rm em})} \,.
\label{tau1}
\end{equation}
When $U^j = V^j$, we obtain immediately that $\tau_{ik}$ is of the form:
\begin{equation}
\tau_{ik} = \frac14 U_{(i}\Delta^j_{k)} \left\{
F_{pq}F_{mn} \frac{\partial C^{pqmn}}{\partial U^j}-
\nabla_l \left[F_{pq}F_{mn} \frac{\partial C^{pqmn}}{\partial (\nabla_l U^j)}
\right]\right\} \,.
\label{tau2}
\end{equation}
Clearly, the energy density scalars and pressure tensors, calculated using the
tetrad and aether paradigms, coincide:
\begin{equation}
U^i\tau_{ik}U^k = 0  \ \Rightarrow W^{(\rm aether)} = W^{(\rm tetrad)} \,,
\label{tau3}
\end{equation}
\begin{equation}
\Delta^{i}_{m}\tau_{ik}\Delta^{k}_{n} = 0  \ \Rightarrow {\cal P}_{mn}^{(\rm
aether)} = {\cal P}_{mn}^{(\rm tetrad)} \,.
\label{tau4}
\end{equation}
Only the flux four-vectors differ:
\begin{equation}
{\cal Q}^h_{(\rm aether)}-{\cal Q}^h_{(\rm tetrad)} = \Delta^{ih}\tau_{ik}U^{k}
=
\frac18 \Delta^{jh} \left\{
F_{pq}F_{mn} \frac{\partial C^{pqmn}}{\partial U^j}-
\nabla_l \left[F_{pq}F_{mn} \frac{\partial C^{pqmn}}{\partial (\nabla_l U^j}
\right]\right\} \,.
\label{tau5}
\end{equation}
Thus, we have to return to the Minkowski-Abraham controversy and to discuss this
difference. Let us consider three model before starting to analyze the problem.

\section{Three examples of the linear response tensor}

\subsection{Axionic vacuum}

In the first example the linear response tensor is assumed to contain neither
velocity four-vector, nor its covariant derivative:
\begin{equation}
C^{pqmn}_{(\rm vacuum)} = \frac12 \left(g^{pqmn} +  \phi \ \epsilon^{pqmn}
\right)
\,. \label{ex1}
\end{equation}
Here and below we use the auxiliary tensor
\begin{equation}
g^{pqmn}\equiv g^{pm} g^{qn}- g^{pn}g^{qm}
\,. \label{ex111}
\end{equation}
Using the definitions (\ref{emunu}) we obtain
\begin{equation}
\varepsilon^{im} = \Delta^{im} \,, \quad (\mu^{-1})_{pq} = \Delta_{pq} \,,
\quad \nu_{p}^{ \ m} = - \phi \Delta_p^m \,. \label{emunu77}
\end{equation}
Since $\nu_{p}^{ \ m} \neq 0$, this medium possesses magnetoelectric properties,
which are provided by the presence of the pseudoscalar field $\phi$.
Calculations in both: tetrad and aether paradigms (see (\ref{em5}) and
(\ref{em52}), respectively), give the same traceless tensor
\begin{equation}
T_{ik}^{(\rm vacuum)} = {\cal T}_{ik}^{(\rm vacuum)} = \frac14 g_{ik} F_{mn}
F^{mn} - F_{im} F_{k}^{\ m}
\,. \label{ex2}
\end{equation}
In other words, the stress-energy tensors do not differ one from another, and
they do not contain axionic field.
Respectively, the energy density scalars, energy flux four-vectors and pressure
tensors
$$
W = -\frac12 \left(E^mE_m + B^mB_m \right) \,, \quad {\cal Q}^j = - \eta^{jmn}
E_m B_n \,,
$$
\begin{equation}
{\cal P}^{pq} = \frac12 \Delta^{pq}\left(E^mE_m + B^mB_m \right) - \left(E^pE^q
+ B^pB^q \right)
\,. \label{ex3}
\end{equation}
formally coincide for both definitions of the velocity four vector, $V^i$ and
$U^i$.

\subsection{Spatially isotropic homogeneous moving dielectric medium}

\subsubsection{Calculations in the context of the tetrad paradigm}

The linear response tensor contains now terms quadratic in the velocity
four-vector:
\begin{equation}
C^{pqmn}=C^{pqmn}_{(0)}+ C^{pqmn}_{(\phi)} \,,
\label{iso1}
\end{equation}
\begin{equation}
C^{pqmn}_{(0)}=\frac{1}{2\mu}\left[\left(g^{pm}g^{qn}{-}g^{pn}g^{qm}\right)+\left(\varepsilon
\mu{-}1\right)\left(g^{pm}V^{q}V^{n}{-}g^{pn}V^{q}V^{m}{+}g^{qn}V^{p}V^{m}{-}g^{qm}V^{p}V^{n}\right)\right]
\,, \label{iso2}
\end{equation}
\begin{equation}
C^{pqmn}_{(\phi)} \equiv  \frac{1}{2}\phi\left[\epsilon^{pqmn}+ \nu g_{rh}V^h
\left(V^{p}\epsilon^{rqmn}-V^{q}\epsilon^{rpmn} +
V^{m}\epsilon^{rnpq}-V^{n}\epsilon^{rmpq}\right) \right]
\,. \label{iso199}
\end{equation}
Using the definitions (\ref{emunu}) we again calculate the permittivity tensors
and the tensor of magneto-electric coefficients:
\begin{equation}
\varepsilon^{im} = \varepsilon \Delta^{im} \,, \quad (\mu^{-1})_{pq} =
\frac{1}{\mu}\Delta_{pq} \,, \quad  \nu_{p}^{ \ m} = - \phi \Delta_p^m (1+\nu)
\,.
\label{emunu88}
\end{equation}
Thus, $\varepsilon$ characterizes the dielectric permittivity;  $\mu$ is the
constant of  magnetic permeability; $n = \sqrt{\varepsilon \mu}$ is the
refraction index; $\nu$ is the magnetoelectric constant. When $\varepsilon=1$,
$\mu=1$, $\nu=0$, the tensor $C^{pqmn}$ converts into $C^{pqmn}_{(\rm vacuum)}$
(\ref{ex1}).
The stress-energy tensor calculated using (\ref{em5}) can be presented in two
forms. The first representation contains the Maxwell tensor:
\begin{equation}
T_{ik}^{({\rm isotropic})}=\frac{1}{4}g_{ik} F_{pq}F_{mn}C^{pqmn}_{(0)}-
\frac12\left[g_{pi} F_{kq} + g_{pk} F_{iq}  \right]C_{(0)}^{pqmn} F_{mn} \,.
\label{iso19}
\end{equation}
The term $C_{(\phi)}^{pqmn}$ disappears from the stress-energy tensor of the
electromagnetic field due to the relations (\ref{tetra9}), and due to the
identity
\begin{equation}
F^{im}F^{*}_{km} = \frac14 \delta^i_k F^{mn} F^{*}_{mn}\,.
\label{iso197}
\end{equation}
The second form of the stress-energy tensor contains the four-vectors $E^i$ and
$B^k$:
$$
T_{ik}^{({\rm isotropic})}=\left(\frac12 g_{ik}-V_i V_k \right)\left(\varepsilon
E^mE_m + \frac{1}{\mu}B^mB_m \right) - \left(\varepsilon E_iE_k +
\frac{1}{\mu}B_iB_k \right) -
$$
\begin{equation}
- \frac12\left(\varepsilon + \frac{1}{\mu} \right) \left(V_i \eta_{kmn}+V_k
\eta_{imn} \right)E^m B^n  \,.
 \label{iso21}
\end{equation}
Clearly, the tensor (\ref{iso21}) is traceless, and it contains neither the
parameter $\nu$, nor the pseudoscalar (axion) field.
The formulas
$$
W = -\frac12 \left(\varepsilon E^mE_m + \frac{1}{\mu}B^mB_m \right) \,, \quad
{\cal Q}^j = - \frac12 \left(\varepsilon + \frac{1}{\mu} \right) \eta^{jmn} E_m
B_n \,,
$$
\begin{equation}
{\cal P}^{pq} = \frac12 \Delta^{pq}\left(\varepsilon E^mE_m +
\frac{1}{\mu}B^mB_m \right) - \left(\varepsilon E^pE^q + \frac{1}{\mu}B^pB^q
\right)
\label{iso22}
\end{equation}
describe the energy density of the electromagnetic field, energy flux
four-vector and pressure tensor, respectively, when $\epsilon \neq 1$, $\mu \neq
1$, $\nu \neq 0$.

\subsubsection{Calculations in the context of the aether paradigm}

Calculations based on the formula (\ref{em52}) yields the following
stress-energy tensor:
$$
{\cal T}_{ik}^{({\rm isotropic})}=\left(\frac12 g_{ik}-U_i U_k
\right)\left(\varepsilon E^mE_m + \frac{1}{\mu}B^mB_m \right) -
\left(\varepsilon E_iE_k + \frac{1}{\mu}B_iB_k \right) -
$$
\begin{equation}
-  \frac{1}{\mu}  \left(U_i \eta_{kmn}+U_k \eta_{imn} \right)E^m B^n  \,.
 \label{2iso21}
\end{equation}
Clearly, the corresponding energy-density scalar and the pressure tensor
\begin{equation}
{\cal W} = -\frac12 \left(\varepsilon E^mE_m + \frac{1}{\mu}B^mB_m \right) \,,
\label{2iso227}
\end{equation}
\begin{equation}
{\cal P}^{pq} = \frac12 \Delta^{pq}\left(\varepsilon E^mE_m +
\frac{1}{\mu}B^mB_m \right) - \left(\varepsilon E^pE^q + \frac{1}{\mu}B^pB^q
\right)
\label{2iso2277}
\end{equation}
coincide with the ones obtained in the framework of the tetrad paradigm.
However, the energy flux four-vector
\begin{equation}
{\cal Q}^j = - \frac{1}{\mu} \eta^{jmn} E_m B_n
\label{2iso228}
\end{equation}
differs from the one given by (\ref{iso22}) by the constant multiplier
$\frac12(n^2+1)$, which is in evident concordance with (\ref{tau5}).

\subsection{Dynamo-optically active medium}

We work in the linear electrodynamics of the chiral (quasi)medium, i.e., adding
a new sophisticated element into the linear response tensor $C^{pqmn}$ we obtain
a new additional term in the corresponding stress-energy tensor. That is why, as
the third example, we consider the model with the linear response tensor, which
is simplified to have the following form in the framework of the tetrad
paradigm:
\begin{equation}
C^{pqmn} = \frac12 g^{pqmn} + X^{lspqmn} g_{js} \nabla_l V^j \,.
\label{dyn1}
\end{equation}
When we deal with the aether paradigm, we have to replace $V^i$ with $U^i$. In
other words, we consider the dynamo-optically active vacuum with $\varepsilon
{=} 1$, $\mu {=} 1$, $\nu {=} 0$.
The new constitutive tensor
\begin{equation}
X^{lspqmn}=\frac{1}{4}g_{ar}V^{r}
g_{bt}V^{t}\left[\alpha\left(g^{pqla}g^{mnsb}+g^{mnla}g^{pqsb}\right)-
\gamma\left(\epsilon^{pqla}\epsilon^{smnb}+\epsilon^{pqsa}\epsilon^{lmnb}\right)\right]
\label{dyn2}
\end{equation}
is assumed to contain two new coupling constants $\alpha$ and $\gamma$ (see
\cite{BL} for the complete representation of this constitutive tensor). In order
to interpret these coupling constants, we calculate the tensors
$\varepsilon^{ik}$, $\left(\mu^{-1}\right)^{ik}$ and
$\nu^{ik}$, and obtain that
\begin{equation}
\varepsilon^{ik} = \Delta^{ik} + \alpha \nablab^{(i} V^{k)} \,, \quad
\left(\mu^{-1}\right)^{ik} = \Delta^{ik} + \gamma \nablab^{(i} V^{k)} \,, \quad
\nu^{ik}=0 \,.
\label{dyn3}
\end{equation}
Thus, the parameter $\alpha$ is associated with the dynamo-optically induced
dielectric susceptibility, while $\gamma$ relates to the dynamo-optically
induced magnetic susceptibility. Now we are ready for calculations of the
stress-energy tensor components.

\subsubsection{Analysis based on the tetrad paradigm}

We use the already obtained tensor (\ref{ex2}) and present the whole
stress-energy tensor in the following tentative form:
$$
T_{ik}^{(\rm dynamo)} - T_{ik}^{(\rm vacuum)} =
$$
$$
=
\frac{1}{4}F_{pq}F_{mn}\left\{ g_{ik} (\nabla_l V_s) \left[X^{lspqmn}  -
\epsilon^{fhrt}\frac{\partial X^{lspqmn}
}{\partial\epsilon^{fhrt}}\right]-\left(\delta_{i}^{l}\delta_{k}^{s}+\delta_{k}^{l}\delta_{i}^{s}\right)
(\nabla_h V^j)\frac{\partial \left(X^{hfpqmn} g_{jf}\right)}{\partial
g^{ls}}-\right.
$$
$$
\left.-\frac{1}{2}\left(V_{i}\delta_{k}^{j}+V_{k}\delta_{i}^{j}\right)
\left(\nabla_l V_s \right) \frac{\partial X^{lspqmn}}{\partial
V^{j}}-\frac{1}{2}\left(g_{is}\nabla_{l}V_{k}+g_{ks}\nabla_{l}V_{i}\right)\
X^{lspqmn} \right\} +
$$
\begin{equation}
+\frac{1}{8}\nabla_h \left\{F_{pq}F_{mn}
\left[\left(V_{i}g_{lk}+V_{k}g_{li}\right) X^{lhpqmn} -\left( g_{is}g_{lk}+
g_{ks}g_{li}\right) V^{h}X^{lspqmn}\right]\right\} \,,
\label{dyn4}
\end{equation}
where the tensor $X^{lspqmn}$ is given by (\ref{dyn2}). Further routine but
cumbersome calculations give the following result:
$$
T_{ik}^{(\rm dynamo)}  =  T_{ik}^{(\rm vacuum)} +
$$
$$
+ \left(\frac12 g_{ik}- V_i V_k \right) \left(\alpha E^l E^s + \gamma B^l B^s
\right)\nablab_{(l}V_{s)} - \frac{1}{2}  \left[\alpha E^lE^s+\gamma B^l
B^s\right] \nablab_{l}V_{(k}g_{i)s} -
$$
$$
- \frac12 \left[\alpha E^s V_{(i}E_{k)} - \gamma B^s V_{(i}B_{k)} \right] DV_s
-\alpha E^j E_{(i} \nablab_{k)} V_j -
$$
$$
-\left[\alpha B^m E^{(s} \eta^{l)}_{\ \ m(k} V_{i)} - \gamma E^m B^{(s}
\eta^{l)}_{\ \ m(k} V_{i)} \right]\nablab_{l}V_{s} +
$$
\begin{equation}
+\frac{1}{2}\nabla_h \left\{\alpha E^h V_{(i}E_{k)} -\gamma B^h V_{(i}B_{k)} -
V^{h} \left[\alpha E_iE_k-\gamma B_i B_k \right] \right\} \,.
\label{dyn5}
\end{equation}
We are interested to find the energy flux four-vector associated with this
tensor; it is now of the following form:
$$
{\cal Q}^h_{(\rm tetrad)} \equiv  \Delta^{hi}T_{ik}^{(\rm dynamo)} V^k =
$$
\begin{equation}
= \eta^{hmn} B_m E_n +
\frac14 \Delta^h_s \nablab_l \left(\alpha E^l E^s - \gamma B^l B^s \right) +
\frac12 \left[\alpha B^m \eta^h_{\ \ m(l}E_{s)} - \gamma E^m \eta^h_{\ \
m(l}B_{s)} \right] \nablab^{(l} V^{s)}
\,.
\label{dyn6}
\end{equation}
Keeping in mind that according to (\ref{act4}) $\nablab^{(l} V^{s)} =
\sigma^{ls} + \frac13 \Theta \Delta^{ls}$, we can say that the energy flux
depends on the shear tensor $\sigma^{ls}$ and on the expansion scalar $\Theta$
of the velocity flow, but it ignores the acceleration and rotation of the
dynamo-optically active medium described by the presented model.

\subsubsection{Analysis based on the aether paradigm}

In order to describe the stress-energy tensor in the framework of the aether
paradigm, we use the consequence of the formulas (\ref{tau1}) and (\ref{tau2}),
which  now can be written as follows:
$$
{\cal T}_{ik}^{(\rm dynamo)} - T_{ik}^{(\rm dynamo)} =
$$
\begin{equation}
- \frac12 U_{(i}\Delta_{k)s} \nablab_{l} \left[\alpha E^l E^s - \gamma B^l B^s
\right] +
\nablab_{l}U_{s} \left[\alpha B^m E^{(s}\eta^{l)}_{\ \ m(k} U_{i)} +
\gamma  E^m B^{(s}\eta^{l)}_{\ \ m(k} U_{i)} \right] \,.
\label{dyn7}
\end{equation}
As it was mentioned above, only the flux four-vectors do not coincide for these
two approaches, giving the following difference:
$$
{\cal Q}^h_{(\rm aether)}-{\cal Q}^h_{(\rm tetrad)} =
$$
\begin{equation}
= -\frac14 \Delta^h_{s}\nablab_{l} \left(\alpha E^l E^s - \gamma B^lB^s \right)
- \frac12 \eta^h_{\ \ ml}\left[\alpha E_{s}B^m + \gamma B_{s} E^m \right]
\nablab^{(l}U^{s)}
\,.
\label{dyn8}
\end{equation}
This final result is
\begin{equation}
{\cal Q}^h_{(\rm aether)}= \eta^{hmn} B_m E_n - \gamma E^m \eta^h_{\ \ m(l}
B_{s)} \nablab^{(l} V^{s)} \,,
\label{dyn99}
\end{equation}
i.e., the energy flux four-vector in the dynamo-optically active medium,
calculated in the approach, which we indicated as aether paradigm, does not
contain the susceptibility parameter $\alpha$, but includes the parameter
$\gamma$.

\section{Discussion}

Readers could ask the authors, what is an expediency to follow sophisticated
calculations presented above? Are there some applications of the developed
formalism? Answering the last question we would like to recall only one fact.
The interpretation of the outstanding astronomical event GW170817 / GRB 170817A
(see \cite{170817}), which  is connected with the discovery of gravitational
waves and gamma-rays from a binary neutron star merger, is based on the standard
model of the electromagnetic wave propagation and the energy transfer. In other
words, for the interpretation of this event the standard formula for the
electromagnetic energy flux in vacuum was used. Let us imagine now, that the
dynamic aether really exists, that this aether is dynamo-optically active, and
that the electromagnetic radiation from the binary system propagates indeed
inside the dynamic aether. Then we have to use the formula
(\ref{dyn99}) for estimations . Since we keep in mind the cosmological context, we consider the aether flow to possess only the expansion, so that the covariant derivative of
the velocity four-vector of the aether has the form $\nabla_i U_k = H(t)
\Delta_{ik}$, where $H(t)=\frac13 \Theta$ is the Hubble function. Then the
formula (\ref{dyn99}) reduces to ${\cal Q}^h_{(\rm aether)}= \eta^{hmn} B_m E_n
[1+ \gamma H(t)]$, and the energy flux four-vector differs from the Poynting
vector by the multiplier $[1+ \gamma H(t)]$. Is it possible to find this
multiplier from observations? It is a not easy question, but certainly it is
very interesting one, and we hope to return to this problem in a special work.

\vspace{8mm}
\noindent
{\bf Acknowledgements}

\noindent
The work was supported by the Program of Competitive Growth of Kazan Federal
University.

\end{document}